\documentclass[amsmath,amssymb,showpacs]{revtex4}

\usepackage{graphicx}
\usepackage{dcolumn}
\usepackage{bm}

\begin{document}

\title{
Effective Sampling in the Configurational Space \\
by the Multicanonical-Multioverlap Algorithm 
} 

\author{
Satoru G. Itoh\footnote{Electronic address: itoh@tb.phys.nagoya-u.ac.jp} 
and Yuko Okamoto\footnote{Electronic address: okamoto@phys.nagoya-u.ac.jp}
}
\affiliation{
Department of Physics \\
School of Science \\
Nagoya University \\ 
Nagoya, Aichi 464-8602, Japan \\
}

\begin{abstract}

We propose a new generalized-ensemble algorithm, which we refer to as the multicanonical-multioverlap algorithm.
By utilizing a non-Boltzmann weight factor, 
this method realizes a random walk in the multi-dimensional, energy-overlap space 
and explores widely in the configurational space including specific configurations, 
where the overlap of a configuration with respect to a reference state is a measure for structural similarity. 
We apply the multicanonical-multioverlap molecular dynamics method to 
a penta peptide, Met-enkephalin, in vacuum as a test system.
We also apply the multicanonical and multioverlap molecular dynamics methods to 
this system for the purpose of comparisons.
We see that the multicanonical-multioverlap molecular dynamics method realizes effective sampling 
in the configurational space including specific configurations more than the other two methods. 
From the results of the multicanonical-multioverlap molecular dynamics simulation, 
furthermore, 
we obtain a new local-minimum state of the Met-enkephalin system. 

\end{abstract}

\maketitle

%

  \section{Introduction} 
  \label{intro:sec}

In order to understand the protein folding mechanisms, 
it is essential that we investigate free-energy landscapes of protein systems. 
In experiments, however, to obtain the free-energy landscapes of protein systems is very difficult. 
Therefore, computer simulations are now widely used for problems of the protein folding mechanisms.
In computer simulations, however, it is still difficult to get accurate free-energy landscapes of protein systems. 
Accordingly, many efforts are devoted to develop efficient simulation algorithms. 

In complex systems such as proteins, 
we must realize effective samplings in the configurational space. 
In usual canonical-ensemble simulations \cite{mrrtt53,hlm82,evans83,nose_mp84,nose_jcp84,hoover85}, however, 
it is difficult to achieve this. 
This is because the usual canonical-ensemble simulations tend to get trapped in a few of many local-minimum states. 
To overcome these difficulties, the generalized-ensemble algorithms have been proposed 
(for a review, see, for instance, Ref.~\cite{mso01}). 

The multicanonical algorithm \cite{berg91,berg92,hoe96,naka97}
is one of the most well-known methods among the generalized-ensemble algorithms. 
In the multicanonical ensemble, 
the probability distribution of the potential energy is expressed by the product of the density of states and 
a non-Boltzmann weight factor, which we refer to as the multicanonical weight factor, 
and we have a flat probability distribution of the potential energy. 
Therefore, multicanonical-ensemble simulations realize free random walks in the potential-energy space 
and have effective samplings in the configurational space. 
The random walk in the potential-energy space allows one to calculate 
various thermodynamic quantities as functions of temperature for a wide temperature range 
from the results of a single simulation run by the single-histogram reweighting techniques \cite{fs88,fs89}.
This method is suitable to sample widely the configurational space, 
but not to have samplings that focus on a specific configuration because of the very nature of the algorithm. 
Consequently, 
it is difficult to obtain accurate free-energy landscape around specific configurations 
in multicanonical simulations, while we can obtain the free-energy landscape over wide areas. 

The multioverlap algorithm was recently proposed for Monte Carlo (MC) method \cite{berg03} 
and molecular dynamics (MD) method \cite{itoh04,itoh06} 
in order to investigate the stability of specific configurations and 
transition states among specific configurations. 
In the multioverlap ensemble, 
the probability distribution is expressed by the product of the density of states and 
a non-Boltzmann weight factor, which we refer to as the multioverlap weight factor, 
and we have a flat probability distribution in the overlap space. 
The method aims at achieving effective samplings that focus on specific configurations. 
Accordingly, 
we can obtain an accurate free-energy landscape around the specific configurations and 
estimate correctly transition states among the specific configurations. 
In this method, however, we do not have a random walk in the potential-energy space, and hence, 
thermodynamic quantities can be obtained at temperatures only near the simulation temperatures.

In this article, we propose a simulation method, 
which we refer to as the multicanonical-multioverlap method, 
to sample widely the configurational space and effectively the vicinity of specific conformations of a protein. 
We apply the multicanonical-multioverlap MD method to Met-enkephalin in vacuum 
and test the effectiveness of the method by comparing the results 
with those of the multicanonical MD method and the multioverlap MD method. 

In Section \ref{algo:sec} we summarize the formulation of the multicanonical-multioverlap algorithms. 
We present the details of the three simulations that we performed and their results in Section \ref{app:sec}. 
Section \ref{conc:sec} is devoted to conclusions. 
  \section{Multicanonical-Multioverlap Algorithms} 
  \label{algo:sec}

In this section we describe the multicanonical-multioverlap algorithms. 
In Sec.~\ref{dis:subsec} we define the dihedral-angle distance \cite{berg03,hmy97} 
which is a complementary quantity of the overlap.
Realizing effective samplings in the energy-overlap space, we introduce 
the multicanonical-multioverlap weight factor in Sec.~\ref{eff:subsec}. 
In Sec.~\ref{eofm:subsec} we present the equations of motion 
to perform multicanonical-multioverlap MD simulations. 
We describe multicanonical-multioverlap MC method with 
the Metropolis criterion \cite{mrrtt53} in Sec.~\ref{mc:subsec}.
In Sec.~\ref{reweight:subsec} we explain the reweighting techniques \cite{fs88,fs89}. 
We can obtain appropriate physical quantities at any temperature in the natural ensemble 
by the reweighting techniques. 

  \subsection{Definition of dihedral-angle distance} 
  \label{dis:subsec}

We introduce a dihedral-angle distance, 
which is a complementary quantity of the overlap, as a reaction coordinate. 
The dihedral-angle distance $d$ with respect to a reference configuration is defined by 
\begin{eqnarray}
d = \frac{1}{n \pi} \sum_{i=1}^{n} d_{a}(v_{i},v_{i}^{0})~.
\label{def_dd}
\end{eqnarray}
Here, $n$ is the total number of dihedral angles, 
$v_{i}$ is the dihedral angle $i$,
and $v_{i}^{0}$ is the dihedral angle $i$ of the reference configuration.
The distance $d_{a}(v_{i},v_{i}^{0})$ between two dihedral angles is given by
\begin{eqnarray}
d_{a}(v_{i},v_{i}^{0}) = {\rm min} (|v_{i} - v_{i}^{0}|,2 \pi - |v_{i} - v_{i}^{0}|)~.
\label{def_dis}
\end{eqnarray}
If $d=0$, from Eq.~(\ref{def_dd}), all dihedral angles are coincident with those of the reference configuration, 
and the two structures are identical. 
The dihedral-angle distance is thus an indicator of 
how similar the conformation is to the reference conformation. 

Sampling the vicinity of the reference configuration is equivalent to sampling around $d=0$. 
In order to realize a sampling that focuses on the reference configuration, therefore, 
we just have to sample the neighborhood of $d=0$. 
We then would like to realize an effective sampling, 
which covers widely the configurational space and focuses on the reference configuration. 
In other words, we want to sample effectively and widely the configurational space including near $d=0$. 

  \subsection{Effective sampling in the energy-overlap space} 
  \label{eff:subsec}

In the case of canonical ensemble at a constant temperature $T_{0}$, 
the probability distribution $P_{c}$ of potential energy $E$ is represented by 
the product of the density of states $n(E)$ and the Boltzmann weight factor $W_{c}(E;T_{0})$: 
\begin{eqnarray}
 P_{c}(E;T_{0}) &=& n(E)W_{c}(E;T_{0}) \nonumber \\
 &=& n(E)e^{-\beta_{0} E}~,
\label{P_cano}
\end{eqnarray}
where $\beta_{0}$ is given by $ \beta_{0} = 1/k_{\rm B}T_{0}$ ($k_{\rm B}$ is the Boltzmann constant). 
In the multicanonical ensemble at a constant temperature $T_{0}$, 
by employing the non-Boltzmann weight factor $W_{muca}(E)$, 
which we refer to as the multicanonical weight factor, 
a uniform probability distribution of potential energy is obtained: 
\begin{eqnarray}
P_{muca}(E) &=& n(E) W_{muca}(E) \nonumber \\
&=& n(E) e^{- \beta_{0} E_{muca}(E)} \nonumber \\
&\equiv& {\rm constant}~,
\label{P_muca}
\end{eqnarray}
where $E_{muca}(E)$ is the multicanonical potential energy. 
The multicanonical weight factor, or the multicanonical potential energy, is not {\it a priori} known and 
has to be determined by short preliminary simulations. 
There exist many methods for the determination of the weight factor
(see, e.g., Refs. \cite{bc92,oka95,wprl01,so00,mso03a,mso03b,hansmann97,so02,tmk03}).
Equation (\ref{P_muca}) implies that
Therefore, multicanonical simulations realize a free random walk in the potential-energy space and 
are able to sample effectively the configurational space. 
In this method, however, it is difficult to sample the vicinity of specific configurations 
because of the very nature of the algorithm. 

The multioverlap algorithm, which is developed by generalizing the multicanonical algorithm, 
is suitable to have samplings that focus on specific configurations. 
In the multioverlap ensemble at a constant temperature $T_{0}$, 
the probability distribution of dihedral-angle distances is defined by
\begin{eqnarray}
P_{muov}(d_{1},\cdots,d_{N}) &=& \int dE~n(E;d_{1},\cdots,d_{N}) W_{muov}(E;d_{1},\cdots,d_{N}) \nonumber \\
&=& \int dE~n(E;d_{1},\cdots,d_{N}) e^{- \beta_{0} E + f(d_{1},\cdots,d_{N})} \nonumber \\
&\equiv& {\rm constant}~,
\label{P_muov_m}
\end{eqnarray}
where $d_{i}$ is the dihedral-angle distance with respect to reference configuration $i~(i=1,\cdots,N)$, 
$n(E;d_{1},\cdots,d_{N})$ is the density of states, 
$W_{muov}(E;d_{1},\cdots,d_{N})$ is the multioverlap weight factor, 
and $f(d_{1},\cdots,d_{N})$ is the ``dimensionless free energy''. 
The dimensionless free energy is not {\it a priori} known and 
has to be determined by short preliminary simulations. 
This method performs a random walk in the $N$-dimensional dihedral-angle-distance space, 
in which the simulation visits the $N$ reference configurations often. 
We can thus obtain accurate information about transition states among these $N$ states. 
Because the multioverlap method does not realize a free random walk in the potential-energy space, 
it is difficult to sample widely the potential-energy space. 

We want simulations to have effective and wide samplings including near $d=0$ in the configurational space. 
Therefore, we consider to carry out a simulation 
that performs a random walk both in the potential-energy space and 
in the dihedral-angle-distance space (energy-overlap space). 
In other words, 
the simulation needs to have a constant probability distribution in the energy-overlap space. 
In analogy with the multicanonical ensemble in Eq.~(\ref{P_muca}) or 
the multioverlap ensemble in Eq.~(\ref{P_muov_m}), 
by employing the non-Boltzmann weight factor $W_{mco}(E,d_{1},\cdots,d_{N})$, 
which we refer to as the multicanonical-multioverlap weight factor, 
a uniform probability distribution with respect to the potential energy and dihedral-angle distances is obtained: 
\begin{eqnarray}
P_{mco}(E,d_{1},\cdots,d_{N}) &=& n(E,d_{1},\cdots,d_{N}) W_{mco}(E,d_{1},\cdots,d_{N}) \nonumber \\
&\equiv& {\rm constant}~.
\label{P_mco_m}
\end{eqnarray}
The multicanonical-multioverlap weight factor is not {\it a priori} known and again
has to be determined by short preliminary simulations. 
The multicanonical-multioverlap weight factor $W_{mco}(E,d_{1},\cdots,d_{N})$  at a constant temperature $T_{0}$ 
can be written as  
\begin{eqnarray}
W_{mco}(E,d_{1},\cdots,d_{N}) = e^{- \beta_{0} E_{mco}(E,d_{1},\cdots,d_{N})}~,
\label{W_mco_m}
\end{eqnarray}
where $E_{mco}(E,d_{1},\cdots,d_{N})$ is the multicanonical-multioverlap potential energy. 
We remark that by definition the multicanonical weight factor in Eq.~(\ref{P_muca}) and 
the multicanonical-multioverlap weight factor in Eq.~(\ref{P_mco_m}) are independent of temperature, 
whereas the multioverlap weight factor in Eq.~(\ref{P_muov_m}) depends on temperature. 

  \subsection{Monte Carlo methods in the multicanonical-multioverlap ensemble} 
  \label{mc:subsec}

Canonical MC simulations are performed with the Metropolis criterion \cite{mrrtt53}. 
In the Metropolis criterion, the transition probability from 
state $x$ with potential energy $E$ to state $x^{\prime}$ with potential energy $E^{\prime}$ 
is given by 
\begin{eqnarray} 
w (x \rightarrow x^{\prime}) = \left\{ 
\begin{array}{ll}
1~, & \mbox{for $\Delta E \leq 0~,$ } \\
{\rm exp}(- \beta_{0} \Delta E )~, & \mbox{for $\Delta E > 0~,$ }
\end{array}
\right.
\label{mc_cano}
\end{eqnarray}
where
\begin{eqnarray} 
\Delta E \equiv E^{\prime}-E~.
\label{deltaE_cano}
\end{eqnarray}

In multicanonical-multioverlap MC simulations, 
the transition probability from 
state $x$ with potential energy $E$ and dihedral-angle distances $d_{1},\cdots,d_{N}$ 
to state $x^{\prime}$ with potential energy $E^{\prime}$ and 
dihedral-angle distances $d^{\prime}_{1},\cdots,d^{\prime}_{N}$ is correspondingly given by 
\begin{eqnarray} 
w (x \rightarrow x^{\prime}) = \left\{ 
\begin{array}{ll}
1~, & \mbox{for $\Delta E_{mco} \leq 0~,$ } \\
{\rm exp}(- \beta_{0} \Delta E_{mco} )~, & \mbox{for $\Delta E_{mco} > 0~,$ }
\end{array}
\right.
\label{mc_mco}
\end{eqnarray}
where $E_{mco}$ is the multicanonical-multioverlap potential energy in Eq.~(\ref{W_mco_m}) and 
\begin{eqnarray} 
\Delta E_{mco} \equiv 
E^{\prime}_{mco}(E^{\prime},d^{\prime}_{1},\cdots,d^{\prime}_{N})-E_{mco}(E,d_{1},\cdots,d_{N})~.
\label{deltaE_mco}
\end{eqnarray}

  \subsection{Equations of motion in the multicanonical-multioverlap ensemble} 
  \label{eofm:subsec}

Solving regular Newton's equations of motion leads to the microcanonical ensemble. 
There are several methods to realize the canonical ensemble by the MD simulation 
(see, for example, Refs. \cite{mrrtt53,hlm82,evans83,nose_mp84,nose_jcp84,hoover85}). 
Here, we just consider one of these methods, 
namely the Gaussian constraint method \cite{hlm82,evans83}
(the realization by other methods is also straightforward). 
In the Gaussian constraint method, the following equations of motion with Gaussian thermostat are solved: 
\begin{eqnarray}
\begin{array}{l}
\displaystyle
{\mbox {\boldmath $\dot q$}}_{i}=\frac{d {\mbox {\boldmath $q$}}_{i}}{dt}=
\frac{{\mbox {\boldmath $p$}}_{i}}{m_{i}}~, \vspace*{2mm} \\
{\mbox {\boldmath $\dot p$}}_{i}={\mbox {\boldmath $F$}}_{i}-\zeta_{c} {\mbox {\boldmath $p$}}_{i}~,
\end{array}
\label{eq_cano}
\end{eqnarray}
where $m_{i}$, ${\mbox {\boldmath $q$}}_{i}$, and ${\mbox {\boldmath $p$}}_{i}$ are the mass, 
coordinate vector, and momentum vector of atom $i$. 
The force ${\mbox {\boldmath $F$}}_{i}$ acting on atom $i$ is given by 
\begin{eqnarray}
{\mbox {\boldmath $F$}}_{i}=-\frac{\partial E}{\partial {\mbox {\boldmath $q$}}_{i}}~,
\label{force_cano}
\end{eqnarray}
where $E$ is the potential energy. 
The coefficient $\zeta_{c}$ is chosen so as to guarantee that the total kinetic energy is constant: 
\begin{eqnarray}
\zeta_{c} &=& \frac{\displaystyle \sum_{i} {\mbox {\boldmath $F$}}_{i} \cdot {\mbox {\boldmath $\dot q$}}_{i}}
{\displaystyle 2 \sum_{i} \frac{{\mbox {\boldmath $p$}}^{2}_{i}}{2 m_{i}}}~.
\label{cons_cano}
\end{eqnarray}

Correspondingly, the molecular dynamics algorithm in the multicanonical-multioverlap ensemble 
naturally follows from Eq.~(\ref{W_mco_m}) (see Refs. \cite{hoe96,naka97} for the case of multicanonical MD). 
The multicanonical-multioverlap MD simulation is carried out by 
solving the following modified equations of motion with Gaussian thermostat: 
\begin{eqnarray}
\begin{array}{l}
\displaystyle
{\mbox {\boldmath $\dot q$}}_{i}=\frac{d {\mbox {\boldmath $q$}}_{i}}{dt}=
\frac{{\mbox {\boldmath $p$}}_{i}}{m_{i}}~, \vspace*{2mm} \\
{\mbox {\boldmath $\dot p$}}_{i}={\mbox {\boldmath $F$}}^{mco}_{i}-\zeta_{mco} {\mbox {\boldmath $p$}}_{i}~.
\end{array}
\label{eq_mco}
\end{eqnarray}
The `force' ${\mbox {\boldmath $F$}}^{mco}_{i}$ acting on atom $i$ is calculated from 
\begin{eqnarray}
{\mbox {\boldmath $F$}}^{mco}_{i} = -\frac{\partial E_{mco}}{\partial {\mbox {\boldmath $q$}}_{i}}~,
\label{force_mco}
\end{eqnarray}
where $E_{mco}$ is the multicanonical-multioverlap potential energy in Eq.~(\ref{W_mco_m}). 
The coefficient $\zeta_{mco}$ is defined by 
\begin{eqnarray}
\zeta_{mco}=\frac{\displaystyle \sum_{i} {\mbox {\boldmath $F$}}^{mco}_{i} 
\cdot {\mbox {\boldmath $\dot q$}}_{i}}
{\displaystyle 2 \sum_{i} \frac{{\mbox {\boldmath $p$}}^{2}_{i}}{2 m_{i}}}~.
\label{cons_mco}
\end{eqnarray}

  \subsection{Reweighting techniques} 
  \label{reweight:subsec}

The results of the multicanonical-multioverlap production run can be analyzed by the reweighting techniques. 
Suppose that we have determined the multicanonical-multioverlap potential energy $E_{mco}$ 
at a constant temperature $T_{0}$ and 
that we have made a production run at this temperature. 
The expectation value of a physical quantity $A$ at any temperature $T$
is calculated from 
\begin{eqnarray}
<A>_{T} = \frac{\displaystyle \sum_{E,d_{1},\cdots,d_{N}} A(E,d_{1},\cdots,d_{N}) 
n(E,d_{1},\cdots,d_{N}) e^{-\beta E}}
{\displaystyle \sum_{E,d_{1},\cdots,d_{N}} n(E,d_{1},\cdots,d_{N})e^{-\beta E}}~,
\label{rew_ev_pre}
\end{eqnarray}
where the best estimate of the density of states is given 
by the single-histogram reweighting techniques \cite{fs88,fs89} 
(see Eq.~(\ref{P_mco_m})):
\begin{eqnarray}
n(E,d_{1},\cdots,d_{N}) = \frac{\displaystyle  N_{mco}(E,d_{1},\cdots,d_{N})}
{\displaystyle W_{mco}(E,d_{1},\cdots,d_{N})}~,
\label{rew_dofs}
\end{eqnarray}
and $N_{mco}(E,d_{1},\cdots,d_{N})$ is the histogram of the probability distribution 
that was obtained by the multicanonical-multioverlap production run. 
By substituting Eqs.~(\ref{W_mco_m}) and (\ref{rew_dofs}) into Eq.~(\ref{rew_ev_pre}), we have 
\begin{eqnarray}
<A>_{T} = \frac{\displaystyle \sum_{E,d_{1},\cdots,d_{N}} A(E,d_{1},\cdots,d_{N}) N_{mco}(E,d_{1},\cdots,d_{N}) 
e^{\beta_{0} E_{mco}(E,d_{1},\cdots,d_{N}) -\beta E}}
{\displaystyle \sum_{E,d_{1},\cdots,d_{N}} N_{mco}(E,d_{1},\cdots,d_{N})e^{\beta_{0} 
E_{mco}(E,d_{1},\cdots,d_{N}) -\beta E}}~.
\label{rew_ev}
\end{eqnarray}

We can also calculate the free energy (or, the potential of mean force) with appropriate reaction coordinates. 
For example, the free energy $F(\xi_{1},\xi_{2};T)$ with reaction coordinates $\xi_{1}$ and $\xi_{2}$ 
at temperature $T$ 
is given by 
\begin{eqnarray}
F(\xi_{1},\xi_{2};T) = -k_{\rm B}T{\rm ln}{P_{c}(\xi_{1},\xi_{2};T)}~,
\label{def_free}
\end{eqnarray}
where $P_{c}(\xi_{1},\xi_{2};T)$ is the reweighted canonical probability distribution of $\xi_{1}$ and $\xi_{2}$ 
and given by (see Eq.~(\ref{rew_ev}))
\begin{eqnarray}
P_{c}(\xi_{1},\xi_{2};T) = \frac{\displaystyle \sum_{E,d_{1},\cdots,d_{N}} 
N_{mco}(\xi_{1},\xi_{2};E,d_{1},\cdots,d_{N}) 
e^{\beta_{0} E_{mco}(E,d_{1},\cdots,d_{N}) -\beta E}}
{\displaystyle \sum_{\xi_{1},\xi_{2},E,d_{1},\cdots,d_{N}} 
N_{mco}(\xi_{1},\xi_{2};E,d_{1},\cdots,d_{N}) e^{\beta_{0} E_{mco}(E,d_{1},\cdots,d_{N}) -\beta E}}~,
\label{rew_pc}
\end{eqnarray}
and $N_{mco}(\xi_{1},\xi_{2};E,d_{1},\cdots,d_{N})$ is the histogram of the probability distribution 
that was obtained from the multicanonical-multioverlap production run. 
  \section{Application to Met-enkephalin in Gas Phase} 
  \label{app:sec}

  \subsection{Computational details} 
  \label{det:subsec}

In order to demonstrate the effectiveness of the multicanonical-multioverlap MD method, 
we compare a multicanonical-multioverlap MD simulation with multicanonical and multioverlap MD simulations. 
We apply the three simulations to the system of Met-enkephalin in vacuum. 
Met-enkephalin is one of the simplest peptides and has the amino-acid sequence Tyr-Gly-Gly-Phe-Met. 
This peptide is often adopted as a test system in biomolecular simulations. 
In our simulations 
the N-terminus and the C-terminus were blocked with the acetyl group and the N-methyl group, respectively. 
This is because we wanted the total charge of the Met-enkephalin system to be neutral. 
The force field that we adopted is the CHARMM param 22 parameter set \cite{p22}. 
Leap-frog algorithm \cite{leap} was employed for the numerical integration and 
the time step was taken to be 0.2 fs. 
The reason for using such a small time step is to perform simulations with high accuracy. 

\begin{figure}
\includegraphics[width=6cm,keepaspectratio]{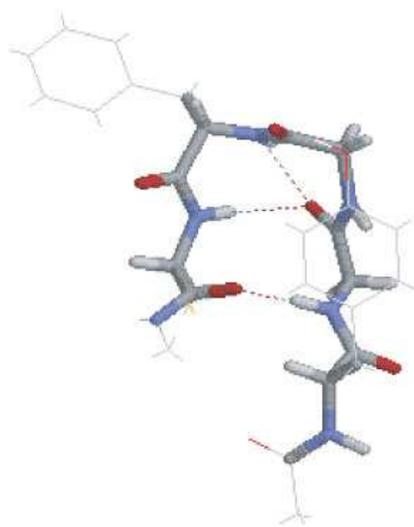}
\caption{
Reference configuration that was used in the multicanonical-multioverlap MD simulation.
The dotted lines denote the hydrogen bonds. 
The N-terminus and the C-terminus are on the right-hand side and on the left-hand side, respectively. 
The figure was created with RasMol \cite{rasmol}. 
}
\label{ref1:fig}
\end{figure}

In the multicanonical-multioverlap MD simulation, 
we must have a reference conformation. 
Therefore, we adopted the conformation in Fig.~\ref{ref1:fig} as the reference conformation and 
set $N=1$ in Eqs.~(\ref{P_mco_m}) and (\ref{W_mco_m}). 
This conformation was obtained by the simulated annealing MD method \cite{kgv83} as follows: 
During the simulated annealing run, 
the temperature was decreased linearly from 1000 K to 100 K with an increment of 50 K, 
and the canonical MD simulations were performed for 200 ps at each temperature. 
This simulated annealing MD run was repeated 10 times with different initial random numbers. 
The obtained final conformations were further minimized by the conjugate gradient method, 
and we finally got two conformations from the results of these minimizations. 
One of the two conformations was the reference conformation in Fig.~\ref{ref1:fig}. 
From our previous results of the multioverlap MD simulations \cite{itoh04,itoh06}, 
we see that this reference conformation is one of the conformations in a local-minimum state and 
another conformation obtained by the results of these minimizations is the global-minimum state 
with the CHARMM param 22 parameter set. 

The backbone dihedral angles are of three types: 
the rotation angle around the $\rm{N-C}^{\alpha}$ bond of the backbone ($\phi$), 
that around the $\rm{C}^{\alpha}-\rm{C}$  bond ($\psi$), 
and that around the peptide bond $\rm{C-N}$ ($\omega$). 
Our multicanonical-multioverlap MD simulation was performed using the all-atom model, 
but we used only $\phi$ and $\psi$ angles in the definition of the dihedral-angle distances in Eq.~(\ref{def_dd}). 
This is because the dihedral angles of the backbone $\omega$ have almost the fixed value of $180^{\circ}$ 
for the peptide bond $\rm{C-N}$. 
Furthermore, by using only the backbone dihedral angles (and not side-chain dihedral angles)
as the elements of the dihedral-angle distances, 
we focused on the backbone structures of Met-enkephalin. 
In Eq.~(\ref{def_dd}), consequently, 
the number $n$ of the elements of the dihedral-angle distances is 10 
because Met-enkephalin has five pairs of $\phi$ and $\psi$.

For the purpose of comparisons, we also performed a multicanonical MD simulation and 
a multioverlap MD simulation for 9 ns at $T_{0}=300~{\rm K}$. 
We determined the multicanonical weight factor in Eq.~(\ref{P_muca}) as follows. 
We carried out canonical MD simulations at eight temperatures between 300 K and 1000 K 
with equal increment of 100 K and 
obtained ensemble-averages of the potential energy at each temperature. 
From the ensemble-averages of the potential energy, 
we calculated the derivative of the multicanonical potential energy \cite{hansmann97,so02,tmk03}: 
\begin{eqnarray}
\left. \frac{\partial E_{muca}(E)}{\partial E} \right|_{E=E_{ave}} = \frac{T_{0}}{T(E_{ave})}~,
\label{W_muca_det1}
\end{eqnarray}
with 
\begin{eqnarray}
E_{ave} = \left<E \right>_{T(E_{ave})}~.
\label{W_muca_det2}
\end{eqnarray}
We integrated the derivative of the multicanonical potential energy and obtained the multicanonical weight factor. 
We adopted a random-coil conformation 
for the initial conformation of the multicanonical MD simulation production run. 

\begin{figure}
\includegraphics[width=6cm,keepaspectratio]{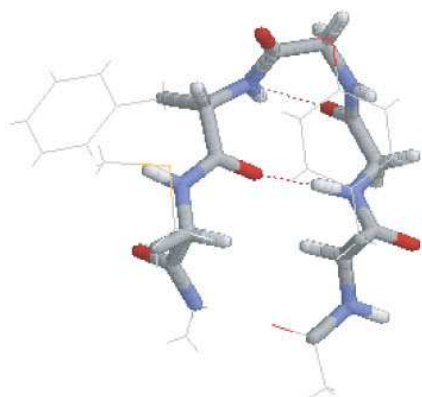}
\caption{
The other reference configuration that was used in the multioverlap MD simulation.
See also the caption of Fig. \ref{ref1:fig}.
}
\label{ref2:fig}
\end{figure}

In the multioverlap MD simulation, we employed the 2-dimensional version of this method. 
In other words, we used two reference conformations in the multioverlap MD simulation and 
$N=2$ in Eq.~(\ref{P_muov_m}). 
One of the two reference conformations was the conformation in Fig.~\ref{ref1:fig}, 
and the other one is shown in Fig.~\ref{ref2:fig}. 
This reference conformation in Fig.~\ref{ref2:fig} is one of the conformations in the global-minimum state 
with the CHARMM param 22 parameter set (see Ref. \cite{itoh04,itoh06}). 
We determined the multioverlap weight factor in Eq.~(\ref{P_muov_m}) by the following 
process \cite{bc92,oka95}. 
Suppose that we have the dimensionless free energy $f=f^{(l)}$ 
in the $l$th iteration of the short multioverlap MD simulation. 
In the $l+1$th iteration of the short multioverlap MD simulation, 
$f^{(l+1)}$ is calculated from 
\begin{equation}
f^{(l+1)}(d_{1},d_{2})=f^{(l)}(d_{1},d_{2})-{\rm log}N^{(l)}(d_{1},d_{2})~,
\label{update_f}
\end{equation}
where $d_{1}$ and $d_{2}$ is the dihedral-angle distance for 
Reference Conformation 1 (RC1) in Fig.~\ref{ref1:fig} and 
Reference Conformation 2 (RC2) in Fig.~\ref{ref2:fig}, respectively. 
$N^{(l)}(d_{1},d_{2})$ in Eq.~(\ref{update_f}) is the histogram obtained from the results of the $l$th iteration. 
For this calculation, the dihedral-angle distances were discretized with a bin size of 0.01. 
Moreover, we interpolated the dimensionless free energy 
by a polynomial, following the techniques that were introduced in Ref.~\cite{okumura06} (see Eq.~(94) there).
The initial value was set as follows: 
\begin{eqnarray}
f^{(1)}(d_{1},d_{2}) = 0~.
\label{init_f}
\end{eqnarray}
In the first iteration, therefore, we performed a short usual canonical MD simulation. 
We stopped the iterations after seven short multioverlap MD simulations each for 1.8 ns and 
we obtained the multioverlap weight factor. 
The initial conformation for the multioverlap production run was a conformation equilibrated at 300 K 
by the canonical simulation. 
Since the multioverlap weight factor has temperature dependence, 
it is appropriate that we employ this initial conformation. 

The multicanonical-multioverlap MD simulation was carried out at $T_{0}=300~{\rm K}$. 
We first have to determine the multicanonical-multioverlap weight factor $W_{mco}(E,d_{1})$ 
in Eq.~(\ref{W_mco_m}) to get a flat probability distribution in the energy-overlap space ($E,d_{1}$). 
For this purpose we used a similar procedure to that in Eq.~(\ref{update_f}). 
Namely, suppose that we have $E_{mco}=E_{mco}^{(l)}$ 
in the $l$th iteration of the short multicanonical-multioverlap MD simulation. 
In the $l+1$th iteration of the short multicanonical-multioverlap MD simulation, 
$E_{mco}^{(l+1)}$ is calculated from 
\begin{equation}
E^{(l+1)}_{mco}(E,d_{1})=E^{(l)}_{mco}(E,d_{1})+k_{\rm B}T_{0}{\rm log}N^{(l)}(E,d_{1})~,
\label{update_E}
\end{equation}
where $N^{(l)}(E,d_{1})$ is the histogram obtained from the results of the $l$th iteration. 
For this calculation, 
the potential energy and the dihedral-angle distance were discretized with 
a bin size of 1.0 kcal/mol and a bin size of 0.01, respectively. 
We also interpolated the multicanonical-multioverlap potential energy by the polynomial. 
The initial value was set as follows: 
\begin{eqnarray}
E_{mco}(E,d_{1}) = E_{muca}(E)~,
\label{first_E}
\end{eqnarray}
where $E_{muca}(E)$ is the multicanonical weight factor 
that was determined as above. 
We then performed three iterations of the multicanonical-multioverlap MD simulations 
in Eq.~(\ref{eq_mco}) for 3 ns. 
The multicanonical-multioverlap weight factor $E_{mco}(E,d_{1})$ was updated by Eq.~(\ref{update_E}) 
after each multicanonical-multioverlap MD simulation. 
Finally, the multicanonical-multioverlap MD production run was then performed with this weight factor for 9 ns 
after equilibration of 1 ns. 
For the initial conformation of the multicanonical-multioverlap MD simulation production run, 
we also adopted a random-coil conformation. 

  \subsection{Comparisons of the three simulations} 
  \label{comp:subsec}

\begin{figure}
\includegraphics[width=13cm,keepaspectratio]{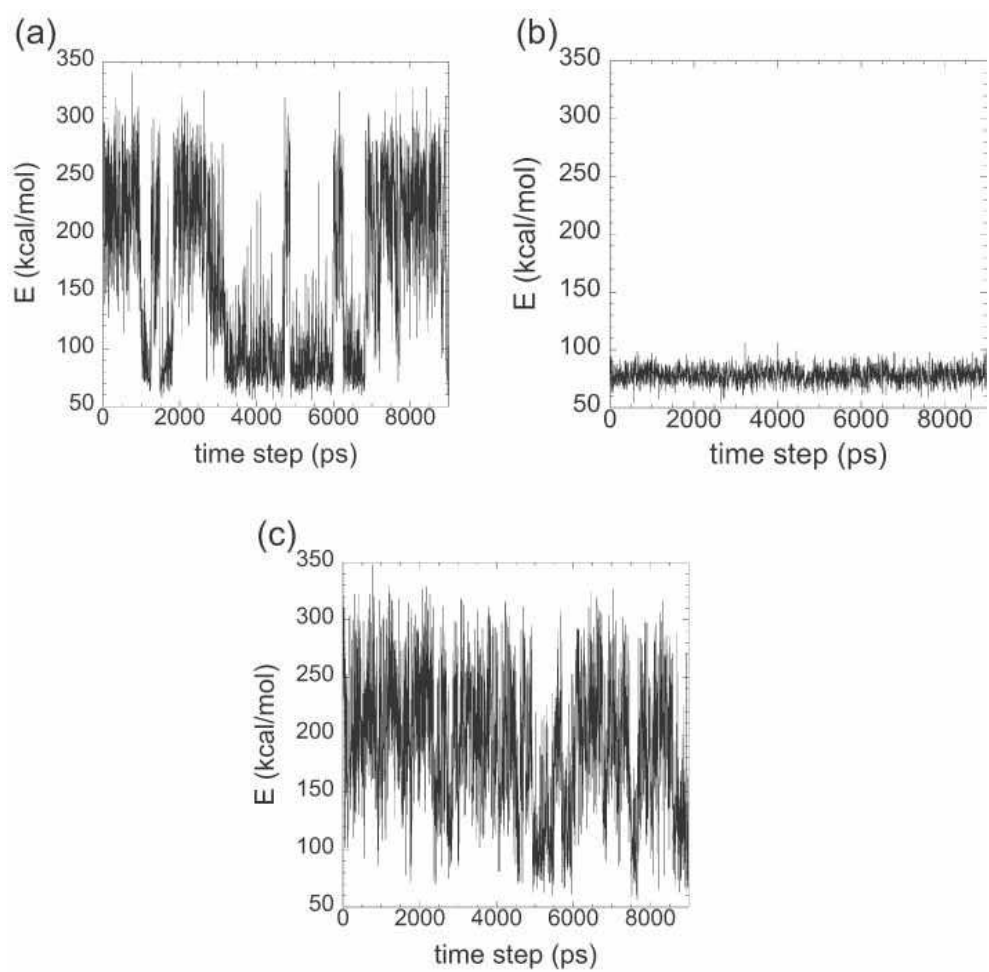}
\caption{
The time series of the potential energy $E$. 
(a) is the results from the multicanonical MD simulation, 
(b) is from the multioverlap MD simulation at $T_{0}=300~{\rm K}$, 
and (c) is from the multicanonical-multioverlap MD simulation.
}
\label{time_E:fig}
\end{figure}

We first compare the time series of the potential energy and the dihedral-angle distance 
obtained from the multicanonical, multioverlap, and multicanonical-multioverlap MD simulations. 
Fig.~\ref{time_E:fig} shows the time series of the potential energy of the three simulations. 
The multicanonical and multicanonical-multioverlap MD simulations cover widely the potential-energy space, 
as we can see in Figs.~\ref{time_E:fig}(a) and \ref{time_E:fig}(c). 
In other words, the two simulations realized free-random walks in the potential-energy space and 
sampled widely the conformational space. 
In the multioverlap MD simulation, however, 
we can sample only a narrow region in the potential-energy space as in Fig.~\ref{time_E:fig}(b). 
Therefore, in contrast with the other two simulations, 
the multioverlap MD simulations are not suitable to sample widely the conformational space. 

\begin{figure}
\includegraphics[width=13cm,keepaspectratio]{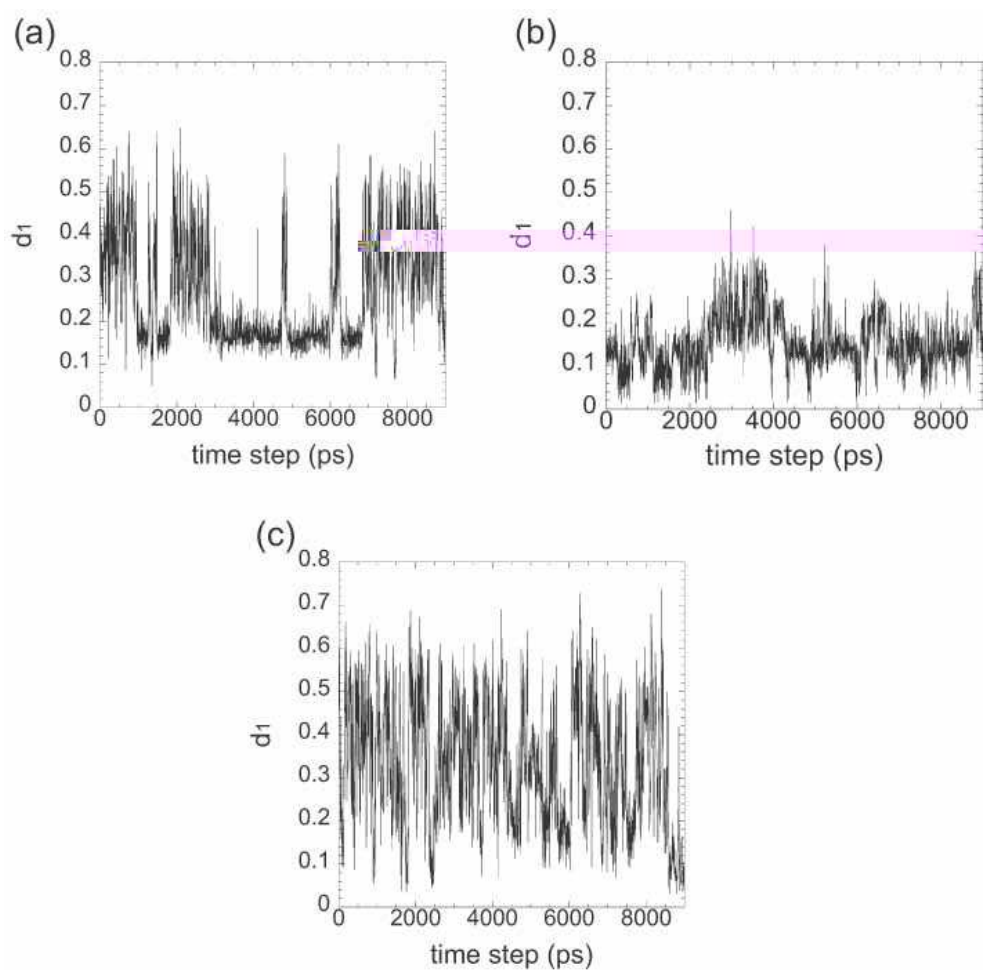}
\caption{
The time series of the dihedral-angle distance $d_{1}$. 
(a) is the results from the multicanonical MD simulation, 
(b) is from the multioverlap MD simulation at $T_{0}=300~{\rm K}$, 
and (c) is from the multicanonical-multioverlap MD simulation.
}
\label{time_d:fig}
\end{figure}

In Fig.~\ref{time_d:fig} we show the time series of the dihedral-angle distance $d_{1}$
with respect to Conformation RC1 in Fig.~\ref{ref1:fig}. 
When $d_{1}=0$, the values of the backbone dihedral angles are completely coincident with those of RC1. 
In the multioverlap and multicanonical-multioverlap MD simulations, 
we see from Figs.~\ref{time_d:fig}(b) and \ref{time_d:fig}(c) 
that the efficient samplings were realized in the neighborhood of $d_{1}=0$. 
In other words, the multioverlap and multicanonical-multioverlap MD simulations could sample efficiently 
the vicinity of RC1. 
The multicanonical MD simulation sampled infrequently the neighborhood of $d_{1}=0$, 
as we can see in Fig.~\ref{time_d:fig}(a). 
Thus, it is difficult to sample specific conformations in multicanonical MD simulations. 

From Fig.~\ref{time_E:fig}(c) and Fig.~\ref{time_d:fig}(c) 
we see that the multicanonical-multioverlap MD simulation sampled both widely in the conformational space and 
efficiently the vicinity of the reference conformation. 
Therefore, the multicanonical-multioverlap MD method has the advantages of 
both the multicanonical MD method and the multioverlap MD method. 

\begin{figure}
\includegraphics[width=13cm,keepaspectratio]{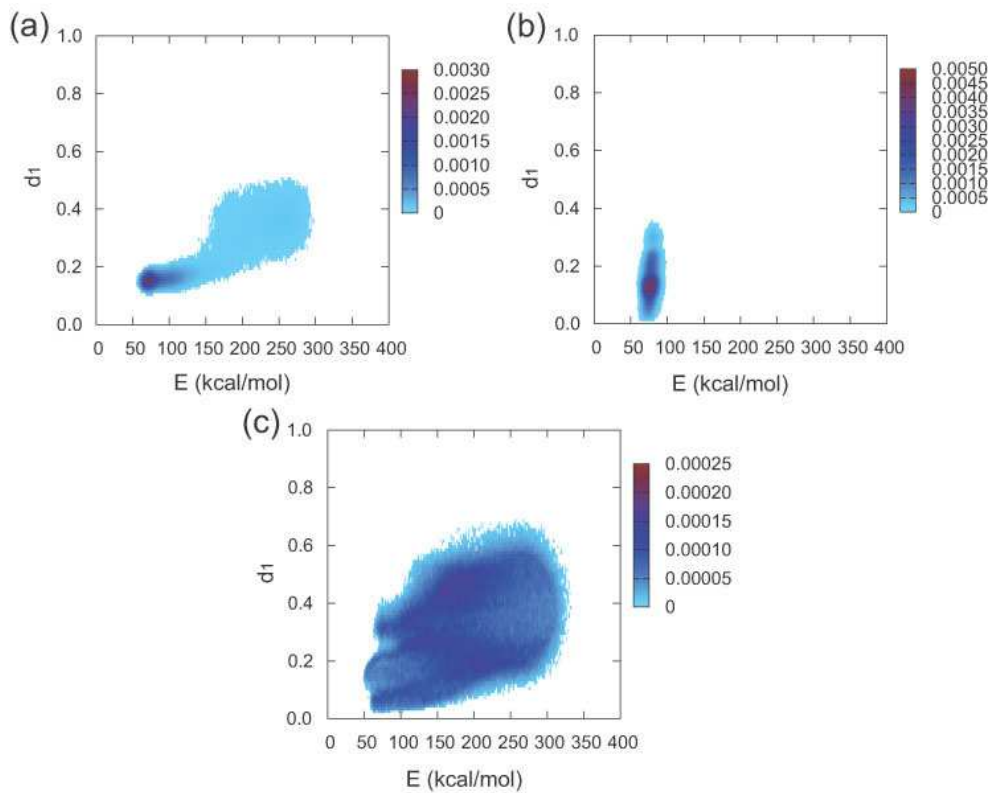}
\caption{
The raw data of the probability distribution with respect to 
the potential energy $E$ and dihedral-angle distance $d_{1}$. 
(a) is the results from the multicanonical MD simulation, 
(b) is from the multioverlap MD simulation at $T_{0}=300~{\rm K}$, 
and (c) is from the multicanonical-multioverlap MD simulation.
}
\label{hist_edc:fig}
\end{figure}

We now consider the probability distributions of configurations from the three simulations. 
In Fig.~\ref{hist_edc:fig} we show the raw data of the histograms 
with respect to the potential energy $E$ and dihedral-angle-distance $d_{1}$. 
From Fig.~\ref{hist_edc:fig}(a) it is apparent that 
the multicanonical MD simulation had only a partial sampling in the vicinity of RC1 (low $d_{1}$ regions). 
Moreover, 
the multicanonical MD simulation did not sample widely in the conformational space at the low-energy region, 
although it had a wide sampling in the conformational space at the high-energy region. 
In the multioverlap MD simulation, on the other hand, 
the sampling that focuses on RC1 was realized (see Fig.~\ref{hist_edc:fig}(b)). 
At the high-energy region, however, we could not sample at all in the multioverlap MD simulation.
From Fig.~\ref{hist_edc:fig}(c) we see that the multicanonical-multioverlap MD simulation performed 
the effective sampling in the conformational space in comparison with the other two simulations. 
In fact, we could sample widely the conformational space at low-energy region as well as high-energy region and 
the vicinity of RC1 in the multicanonical-multioverlap MD simulation. 

  \subsection{Physical quantities calculated by the reweighting techniques} 
  \label{phys:subsec}

\begin{figure}
\includegraphics[width=13cm,keepaspectratio]{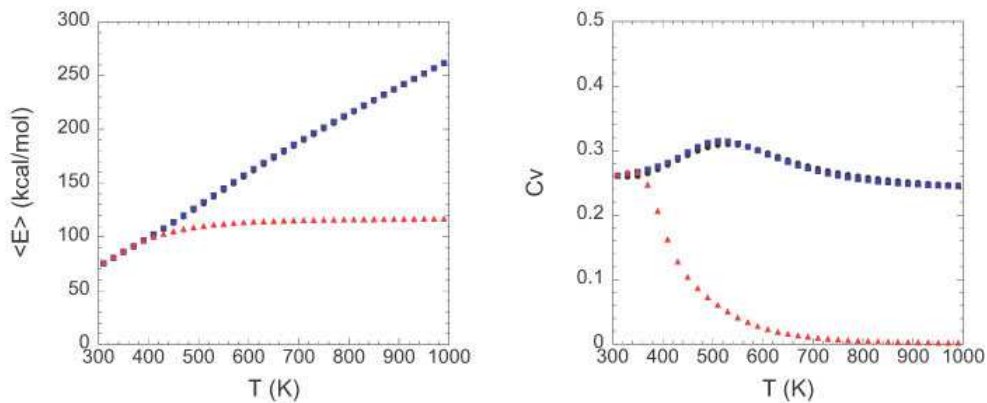}
\caption{
(a) Average potential energy as a function of temperature, and (b) specific heat as a function of temperature. 
These results were 
calculated from the multicanonical MD simulation (blue square), the multioverlap MD simulation (red triangle), 
and the multicanonical-multioverlap MD simulation (black circle) by the reweighting techniques. 
The results from the multicanonical and multicanonical-multioverlap simulations are essentially identical. 
}
\label{rew:fig}
\end{figure}

We present various physical quantities calculated from the results of the three simulations 
by the reweighting techniques. 
The reweighting techniques for the multicanonical-multioverlap MD method 
were explained in Sec.~\ref{reweight:subsec}. 
The reweighting techniques for the other MD methods are accounted, for instance, in Refs.~\cite{mso01,itoh04}. 
In Fig.~\ref{rew:fig} we show the average potential energy and specific heat calculated 
as functions of temperature by the reweighting techniques. 
The specific heat here is defined by 
\begin{eqnarray}
C_{v} &=& \frac{1}{k_{\rm B}} \frac{d \left< E \right>_{T}}{dT} \nonumber \\
&=& \beta^{2} \left( \left< E^{2} \right>_{T} -\left< E \right>^{2}_{T} \right)~.
\label{cv_def:4}
\end{eqnarray}

From the Figure we see that the results from the multicanonical-multioverlap MD simulations 
well coincide with those from the multicanonical MD simulation. 
The results from the multioverlap MD simulation, however, are in agreement with 
those from the other two simulations only near $T=300~{\rm K}$. 
This is because the multioverlap method sampled only a narrow potential-energy range 
as shown in Fig.~\ref{time_E:fig}(b).
Accordingly, we  can obtain various physical quantities at any temperature 
in the multicanonical-multioverlap MD method as in the multicanonical MD method, 
although it is difficult to get such quantities at any temperature in the multioverlap MD method. 

\begin{figure}
\includegraphics[width=13cm,keepaspectratio]{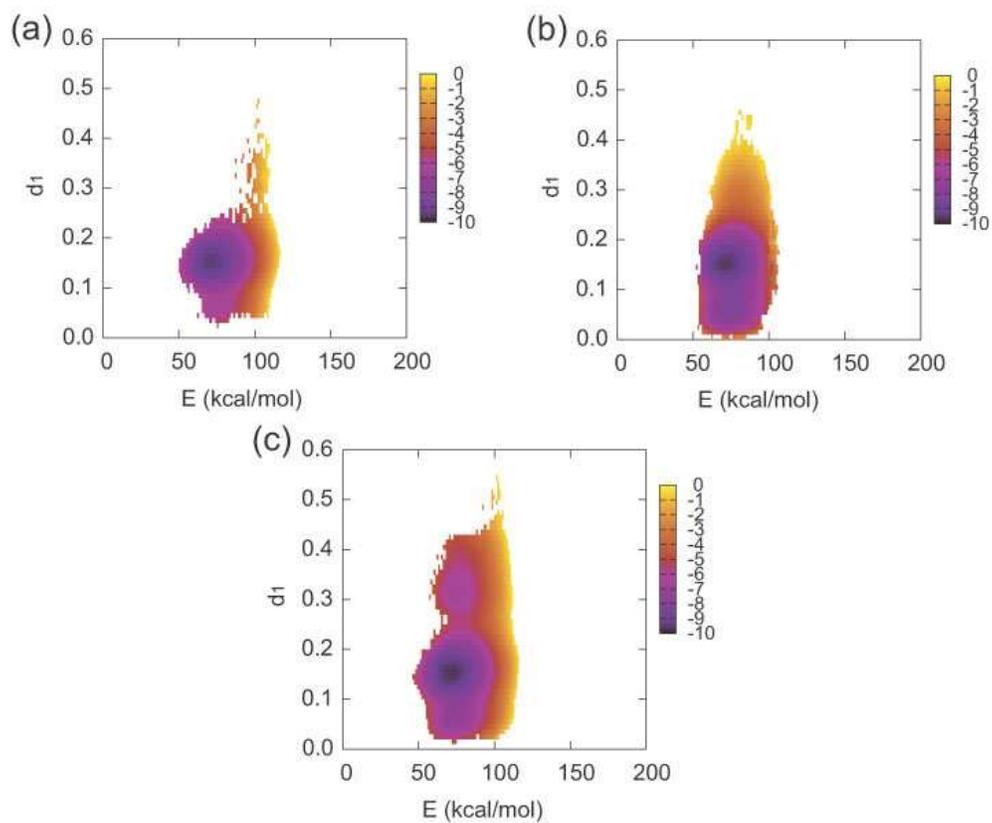}
\caption{
The free-energy landscapes with respect to the potential energy $E$ and dihedral-angle distance $d_{1}$
that were obtained from (a) the multicanonical MD simulation, 
(b) the multioverlap MD simulation at $T_{0}=300~\rm{K}$, 
and (c) the multicanonical-multioverlap MD simulation. 
}
\label{f_ed:fig}
\end{figure}

Subsequently, we present the free-energy landscape with respect to various reaction coordinates. 
The free-energy landscape was calculated from Eq.~(\ref{def_free}) with appropriate reaction coordinates 
by the reweighting techniques. 
In Fig.~\ref{f_ed:fig} 
we show the free-energy landscape at $T=300~{\rm K}$ obtained from the three simulations 
with respect to the potential energy $E$ and dihedral-angle distance $d_{1}$. 
Since the multicanonical MD simulation did not sample the vicinity of RC1 and 
widely the conformational space at the low-energy region, 
the free-energy landscape at $T=300~{\rm K}$ was obtained only in a narrow region 
away from RC1 as shown in Fig.~\ref{f_ed:fig}(a). 
In the multioverlap MD simulation, on the other hand, 
the free-energy landscape in Fig.~\ref{f_ed:fig}(b) including the neighborhood of RC1 was calculated. 
From Fig.~\ref{f_ed:fig}(c), 
the free-energy landscape obtained from the multicanonical-multioverlap MD simulation was described 
over a wide range and near RC1. 
Moreover, we could identify a local-minimum state, 
which is located around $(E,d_{1})=(75.0,0.32)$ in Fig.~\ref{f_ed:fig}(c), 
although we did not find it by the other two methods. 
This is because the multicanonical-multioverlap MD simulation samples widely 
the conformational space at low-energy region and effectively in the vicinity of RC1. 

\begin{figure}
\includegraphics[width=13cm,keepaspectratio]{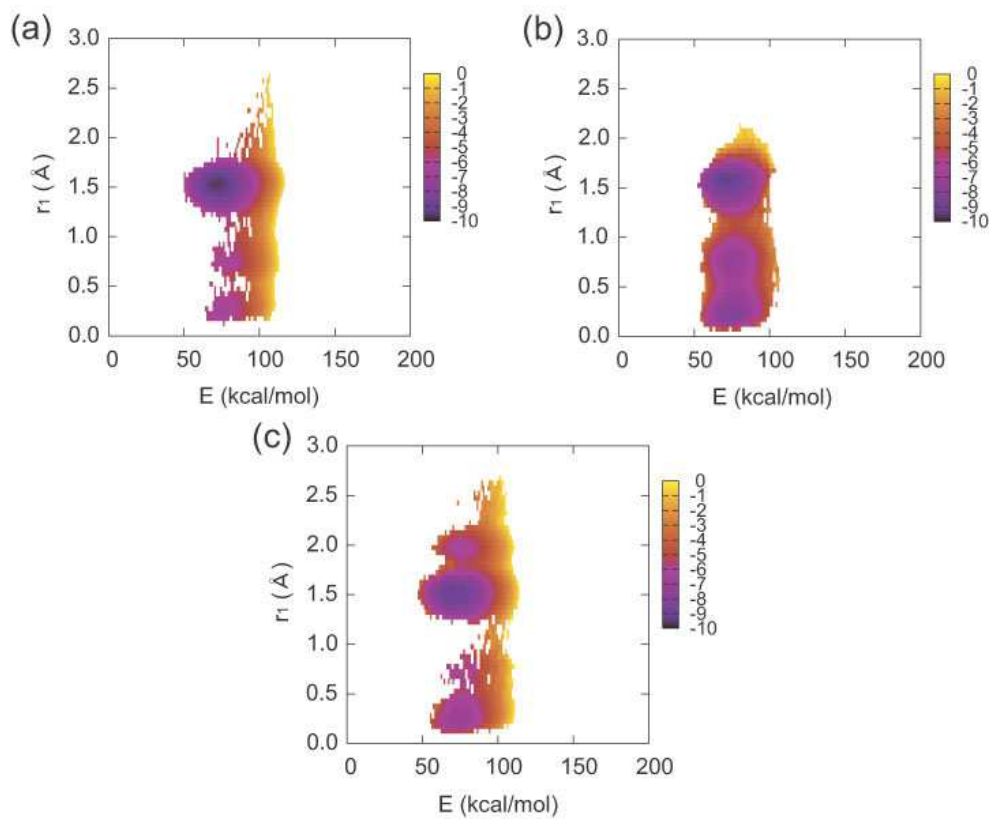}
\caption{
The free-energy landscapes with respect to the potential energy $E$ and RMSD $r_{1}$
that were obtained from (a) the multicanonical MD simulation, 
(b) the multioverlap MD simulation at $T_{0}=300~\rm{K}$, 
and (c) the multicanonical-multioverlap MD simulation. 
}
\label{f_er:fig}
\end{figure}

\begin{figure}
\includegraphics[width=13cm,keepaspectratio]{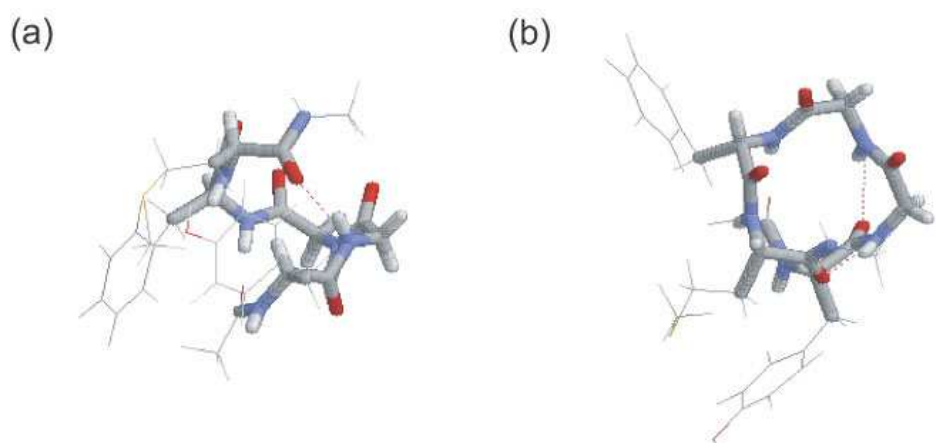}
\caption{
(a) A typical structure in the new local-minimum state found in Figs.~\ref{f_ed:fig}(c) and \ref{f_er:fig}(c).
(a) and (b) correspond to the same conformation viewed from different angles. 
See also the caption of Fig. \ref{ref1:fig}.
}
\label{local:fig}
\end{figure}

In Fig.~\ref{f_er:fig} 
we also show the free-energy landscape at $T=300~{\rm K}$ obtained from the three simulations 
with respect to the potential energy and root-mean-square distance (RMSD) $r_{1}$. 
Here, the RMSD $r_{1}$ with respect to RC1 is defined by 
\begin{eqnarray}
r_{1} = {\rm min} \left[\sqrt{\displaystyle \frac{1}{N} \sum_{i} 
({\mbox {\boldmath $q$}}_{i}-{\mbox {\boldmath $q$}}^{0}_{i})^{2}} \right]~,
\label{def_rmsd}
\end{eqnarray}
where $N$ is the number of atoms, 
$\{ {\mbox {\boldmath $q$}}^{0}_{i} \}$ are the coordinates of RC1, 
and the minimization is taken over the rigid translations and rigid rotations of 
the coordinates of the configuration $\{ {\mbox {\boldmath $q$}}_{i} \}$. 
In this article, we took into account only the backbone coordinates of Met-enkephalin in Eq.~(\ref{def_rmsd}). 
In Fig.~\ref{f_er:fig} 
we can also see a local-minimum state, which is located around $(E,r_{1})=(75.0,2.0)$, 
only from the results of the multicanonical-multioverlap MD simulation. 
By analyzing the conformations in the two local minima found in Figs.~\ref{f_ed:fig}(c) and \ref{f_er:fig}(c), 
we believe that they correspond to the same state. 
This is because common conformations were found in both states. 
In Fig.~\ref{local:fig} we show a representative conformation in this newly found local-minimum state. 
This conformation has a backbone hydrogen bond 
between hydrogen bond donor NH of Gly-2 and hydrogen bond acceptor CO of Met-5.
The conformation also has a backbone hydrogen bond 
between CO of Try-1 and NH of Gly-3 but this hydrogen bond was frequently broken. 
  \section{Conclusions} 
  \label{conc:sec}

In this article, we have proposed the multicanonical-multioverlap MD algorithm, which is useful to 
sample the conformational space widely and the vicinity of a reference conformation effectively. 
We applied this method to a penta-peptide system of Met-enkephalin in vacuum and 
compared the performance with the those of multicanonical and multioverlap MD methods. 
We showed the effectiveness of the multicanonical-multioverlap MD method 
over the multicanonical and multioverlap MD methods. 
The multicanonical MD simulation sampled widely the conformational space at high-energy region 
but not at low-energy region and did not have the sampling around the reference conformation. 
The multioverlap MD simulation could sample effectively the vicinity of the reference conformation. 
In the multioverlap MD simulation, however, we were not able to have the sampling in the high-energy region. 
On the other hand, 
the multicanonical-multioverlap MD simulation realized a free-random walk in the energy-overlap space 
and sampled the conformational space widely and the neighborhood of the reference conformation. 
Accordingly, we could obtain accurate free-energy landscape in the wide reaction-coordinate space 
including the vicinity of the reference conformation and discover a new local-minimum state. 

In the protein folding problem, the multicanonical-multioverlap method can be applied 
to deduce folding pathways in which the protein system has an intermediate state like a molten-globule state. 
This is because we can obtain free-energy landscapes, which include random-coil states, 
the native state, and the molten-globule state, from the results of multicanonical-multioverlap simulations. 
Furthermore, we can estimate transition states accurately 
between the native state (or denatured state) and the molten-globule state 
by employing the molten globule state as the reference conformation in multicanonical-multioverlap simulations. 

\section*{ACKNOWLEDGMENTS}

The computations were performed on the computers at 
the Research Center for Computational Science, Institute for Molecular Science. 
This work was supported, in part, 
by the Grants-in-Aid for the Next Generation Super Computing Project, Nanoscience Program and 
for Scientific Research in Priority Areas, ``Water and Biomolecules'', 
from the Ministry of Education, Culture, Sports, Science and Technology, Japan. 

%
%
%
%
%
%
%
%
%
%
%
%
%
%
%
%
%
\end{document}